# Magnetic Dipole Interaction and Total Magnetic Energy of Lithium Ferrire Thin Films

P. Samarasekara and C. K.D. Sirimanna
Department of Physics, University of Peradeniya, Peradeniya, Sri Lanka

## Abstract
The total magnetic energy of Lithium ferrite thin films was determined using the classical Heisenberg Hamiltonian. The short range magnetic dipole interactions between spins within one unit cell and the interactions between spins in two adjacent unit cells have been determined in order to find the total magnetic energy of lithium ferrite films. Only the spin pairs with separation less than cell constant have been taken into account to calculate dipole interaction and spin exchange interaction. Theoretically several easy and hard directions of lithium ferrite film were found for one set of energy parameters included in our modified Heisenberg Hamiltonian. The dependence of total magnetic energy of a lithium ferrite film on number of unit cells, spin exchange interaction, dipole interaction, second order magnetic anisotropy, fourth order magnetic anisotropy, internal applied magnetic field and stress induced magnetic anisotropy has been explained in this manuscript.

## 1. Introduction:

Lithium Ferrite ($Li_{0.5}Fe_{2.5}O_4$) is a potential candidate in applications of magnetic memory devices, monolithic microwave integrated circuits and microwave devices. The magnetic properties of films depend on the stress of the films induced within cooling and heating process. The crystal structure of lithium ferrite has been explained previously [1]. Magnetic properties of lithium ferrite nanoparticles with a core/shell structure have been investigated [2]. Infrared spectral properties of Magnesium and Aluminum co-substituted Lithium ferrites have been studied [3]. Influence of substrate on the octahedral site order of Structural and magnetic properties of Lithium ferrite thin films have been investigated [4]. The effect of Dipole and Anisotropic Hyperfine Fields on NMR of Fe in Lithium Ferrite has been studied [5].

The short range magnetic dipole interaction of Lithium ferrite thin films have been explained in this manuscript. Previously the magnetic dipole interactions of barium ferrite [6] films have been described by us. However, cobalt belongs to the ferromagnetic category. In all above cases, the total magnetic energy was also determined using the classical Heisenberg Hamiltonian modified by introducing fourth order magnetic anisotropy, stress induced anisotropy and demagnetization factor. In addition, spin exchange interaction, magnetic dipole interaction, second order magnetic anisotropy constant and applied magnetic field terms can be found in this classical Heisenberg Hamiltonian. The investigations of barium ferrite films were restricted to unperturbed Heisenberg Hamiltonian. However, the studies of ferrite and ferromagnetic films were extended to third order perturbed Heisenberg Hamiltonian [7-15]. In addition, the total magnetic energy of cobalt film was determined using second order perturbed Heisenberg Hamiltonian [15]. The technique used to calculate magnetic dipole interaction in our previous reports was employed to find the magnetic dipole interaction of Lithium ferrite films explained in this manuscript. Only the interactions between iron ions were taken into account, since the net magnetic moment of each lithium and oxygen ions is zero.

## 2. Model:

The total energy of a magnetic thin film is given by following modified Heisenberg Hamiltonian [6.]

$$H = -J\sum_{m,n}\vec{S}_m.\vec{S}_n + \omega\sum_{m\neq n}\left(\frac{\vec{S}_m.\vec{S}_n}{r_{mn}^3} - \frac{3(\vec{S}_m.\vec{r}_{mn})(\vec{r}_{mn}.\vec{S}_n)}{r_{mn}^5}\right) - \sum_m D_{\lambda_m}^{(2)}(S_m^z)^2 - \sum_m D_{\lambda_m}^{(4)}(S_m^z)^4$$

$$-\sum_m \vec{H}..\vec{S}_m - \sum_m K_s Sin^2\theta_m \qquad (1)$$

Here $J$, $\omega$, $\theta$, $D_m^{(2)}$, $D_m^{(4)}$, $H_{in}$, $H_{out}$, $K_s$, $m$ and $n$ represent spin exchange interaction, strength of long range dipole interaction, azimuthal angle of spin, second and fourth order anisotropy constants, in plane and out of plane internal magnetic fields, stress induced anisotropy constant and spin plane indices, respectively. When the stress applies normal to the film plane, the angle between $m^{th}$ spin and the stress is $\theta_m$. Therefore the ground state energy will be calculated per spin with Z-axis normal to film plane. The total magnetic energy per unit spin of $Fe^{3+}$ will be determined. The demagnetization factor was not considered in this study. In Lithium ferrite structure, only the $Fe^{3+}$ ions contribute to magnetic moment. Due to the unavailability of unpaired electrons, $Li^+$ ions don't have any net magnetic moment. $Fe^{3+}$ ions occupy octahedral and tetrahedral sites of the lattice as given in table 1.

| Layer | Atomic number | Coordinates | | |
|---|---|---|---|---|
| | | x | y | z |
| A1 | A1$_1$ | 0 | 0 | 0 |
| | A1$_2$ | 1 | 0 | 0 |
| | A1$_3$ | 0 | 1 | 0 |
| | A1$_4$ | 1 | 1 | 0 |
| | A1$_5$ | 0.5 | 0.5 | 0 |
| B1 | B1$_1$ | 0.625 | 0.125 | 0.125 |
| | B1$_2$ | 0.375 | 0.875 | 0.125 |
| | B1$_3$ | 0.125 | 0.625 | 0.125 |
| A2 | A2$_1$ | 0.25 | 0.25 | 0.25 |
| | A2$_2$ | 0.75 | 0.75 | 0.25 |
| B2 | B2$_1$ | 0.625 | 0.375 | 0.375 |
| | B2$_2$ | 0.375 | 0.625 | 0.375 |
| | B2$_3$ | 0.875 | 0.125 | 0.375 |
| A3 | A3$_1$ | 0.5 | 0 | 0.5 |
| | A3$_2$ | 0 | 0.5 | 0.5 |
| | A3$_3$ | 1 | 0.5 | 0.5 |
| | A3$_4$ | 0.5 | 1 | 0.5 |
| B3 | B3$_1$ | 0.125 | 0.125 | 0.625 |
| | B3$_2$ | 0.875 | 0.875 | 0.625 |
| | B3$_3$ | 0.375 | 0.375 | 0.625 |
| A4 | A4$_1$ | 0.75 | 0.25 | 0.75 |
| | A4$_2$ | 0.25 | 0.75 | 0.75 |
| B4 | B4$_1$ | 0.625 | 0.875 | 0.875 |
| | B4$_2$ | 0.875 | 0.625 | 0.875 |
| | B4$_3$ | 0.125 | 0.375 | 0.875 |
| A5 | A5$_1$ | 0 | 0 | 1 |
| | A5$_2$ | 1 | 0 | 1 |
| | A5$_3$ | 0 | 1 | 1 |
| | A5$_4$ | 1 | 1 | 1 |
| | A5$_5$ | 0.5 | 0.5 | 1 |

Table 1: Coordinates of Fe$_{3+}$ ions in a unit cell corresponding to A (tetrahedral) and B (octahedral) sites.

First the spin exchange interactions between spins were determined as following. Only the interactions between spins separated by a distance less than the cell constant were considered. Spin exchange interactions were calculated for nearest spin neighbors within one cell and two adjacent cells as given in table 2.

|     | Interaction | Number of nearest neighbors | |
| --- | --- | --- | --- |
|     |             | Within unit cell | Between adjacent cells |
| $N_A$ | A-A | 70 | 12 |
| $N_B$ | B-B | 63 | 47 |
| $N_{AB}$ | A-B | 230 | 63 |

Table 2: Number of nearest neighbors for each type of interaction.

By using the values given in table 2 and the first term in equation 1, the exchange interaction energy within a unit cell was found as 87J. Similarly the exchange interaction energy between two adjacent unit cells can be expressed as 4J. If there are N number of unit cells in the Lithium ferrite film in z direction, there are (N-1) adjacent cell combinations within the film. Therefore, the total exchange interaction energy can be written as 87NJ+4(N-1)J.

The magnetic dipole interaction energy between nearest spins ($S_i$ and $S_j$) was determined using following equation and matrix.

$$E = \omega \vec{S}_i . W(r_{ij}) . \vec{S}_j \qquad (2)$$

Here 
$$W(r) = \frac{1}{r^3} \begin{pmatrix} 1-3\hat{r}_x^2 & -3\hat{r}_y\hat{r}_x & -3\hat{r}_z\hat{r}_x \\ -3\hat{r}_x\hat{r}_y & 1-3\hat{r}_y^2 & -3\hat{r}_z\hat{r}_y \\ -3\hat{r}_x\hat{r}_z & -3\hat{r}_y\hat{r}_z & 1-3\hat{r}_z^2 \end{pmatrix} \qquad (3)$$

and $\omega = \dfrac{\mu_0 \mu^2}{4\pi a^3}$

Only the interactions between spins separated by a distance less than the cell constant were considered. A film with (001) orientation of Lithium ferrite cell has been considered. As an example, the calculations of some elements of matrix appeared in equation 3 are given below. In layer A1 (bottom layer of the unit cell), five Ferric ions occupy $A1_1$(0, 0, 0), $A1_2$(1, 0, 0), $A1_3$(0, 1,0), $A1_4$(1, 1, 0) and $A1_5$(0.5, 0.5, 0) sites [3]. Because the interactions between spins with separation less than the cell constant have been considered, only the $A1_1A1_5$, $A1_2A1_5$, $A1_3A1_5$ and $A1_4A1_5$ spin interactions have been taken into account.

The spins of Fe ions in the second spin layer located at a distance "a/8" above the bottom layer of Lithium ferrite unit cell occupy $B1_1$(0.125, 0.625, 0.125), $B1_2$(0.375, 0.875, 0.125) and $B1_3$(0.625, 0.125, 0.125) sites [3]. Because of the separations between following interactions are less than the cell-constant, the spin interactions between $A1_1B1_1$, $A1_1B1_2$, $A1_1B1_3$, $A1_2B1_1$, $A1_3B1_2$, $A1_3B1_3$, $A1_4B1_1$, $A1_4B1_2$, $A1_4B1_3$, $A1_5B1_1$, $A1_5B1_2$ and $A1_5B1_3$ were considered for these calculations.

In a similar method, all the matrix elements for adjacent spin layers were calculated, and the dipole interactions for next nearest layers in the cell were calculated. All $W_{11}$, $W_{12}$, $W_{13}$, $W_{21}$, $W_{22}$, $W_{23}$, $W_{31}$, $W_{32}$ and $W_{33}$ matrix elements were calculated from equation 3 using the same method described in one of our previous publications [8]. This matrix was found to be symmetric.

After substituting these matrix elements in the matrix given in equation (3), the dipole matrix can be obtained. By substituting that dipole matrix in equation (2), the dipole interaction energy in each case can be found as following.
The total dipole moment for the spins within A layer can be given as

$E_{dipoleA}$ = -11.31371 $\omega$ (0.25+0.75cos2 $\theta$ )

Similarly for all B type and A-B nearest neighbor interactions within the unit cell, the total dipole interaction energy can be given as

$E_{dipoleB}$ = 20.174631 $\omega$ sin2 $\theta$
$E_{dipoleAB}$ = 26.44766 $\omega$ sin2 $\theta$

Finally the total dipole interaction energy within the unit cell can be given as

$E_{dipole-unit\ cell}$ = $\omega$ (46.52229sin2 $\theta$ −2.82843−8.48528cos2 $\theta$ )          **(4)**

Similarly for all A type, B type and A-B nearest neighbor interactions between two adjacent unit cells in z direction, the total dipole interaction energy can be given as

$E_{dipoleA-adj}$ = -26.06992 $\omega$ (0.25+0.75cos2 $\theta$ )
$E_{dipoleB-adj}$ = - $\omega$ (20.60867+ 22.42482sin2 $\theta$ +66.26353cos2 $\theta$ )
$E_{dipoleAB-adj}$ = $\omega$ (60.1526-25.85484sin2 $\theta$ +180.45779cos2 $\theta$ )

Therefore the total dipole interaction energy between two adjacent unit cells in z direction can be given as

$E_{dipole-adj-cells}$ = $\omega$ (33.02645-48.28966sin2 $\theta$ +94.64182cos2 $\theta$ )          **(5)**

If there is N number of unit cells in the Lithium ferrite film in z direction, then the total dipole interaction energy can be given as

$E_{dipole-total}$ = N$E_{dipoleunit-cell}$ + (N-1)$E_{dipole-adj-cells}$          **(6)**

By substituting equations (4) and (5) in equation (6), the total dipole interaction energy of a Lithium ferrite film with thickness Na (height of N cubic unit cells) in z direction can be given as,
$E_{dipole-total}$ = N $\omega$ (46.62229sin2 $\theta$ −2.82843 −8.48528cos2 $\theta$ )
          +(N-1) $\omega$ (33.02645-48.28966sin2 $\theta$ +94.64182cos2 $\theta$ )          **(7)**

Then by substituting equation (7) in equation (1), the total energy per unit spin can be given as,

E(θ)=(91N-4)J+ω{N(46.62229sin2θ-2.82843-8.48528cos2θ)

+(N-1)(33.02645-48.28996sin2θ+94.64182cos2θ)}

$$-\cos^2\theta\sum_{m=1}^{N}D_m^{(2)}-\cos^4\theta\sum_{m=1}^{N}D_m^{(4)}+4N(H_{in}\sin\theta+H_{out}\cos\theta+K_s\sin 2\theta)$$

Hence

$$\frac{E(\theta)}{\omega}=(91N-4)\frac{J}{\omega}+\text{N}(46.62229\sin2\theta\text{-}2.82843\text{-}8.48528\cos2\theta)$$

+(N-1)(33.02645-48.28996sin2θ+94.64182cos2θ)

$$-\cos^2\theta\sum_{m=1}^{N}D_m^{(2)}-\cos^4\theta\sum_{m=1}^{N}D_m^{(4)}+4N(H_{in}\sin\theta+H_{out}\cos\theta+K_s\sin 2\theta)$$

Hence Following 2-D and 3-D graphs have been plotted using above final energy equation.

## 3. Results and discussion:

The 3-D graph of $\frac{E(\theta)}{\omega}$ versus $\frac{\sum_{m=1}^{N}D_m^{(2)}}{\omega}$ and θ is given in figure 1. The variation of energy is most likely periodic. Values of other parameters were kept at N=1000,

$$\frac{J}{\omega}=\frac{H_{in}}{\omega}=\frac{H_{out}}{\omega}=\frac{K_s}{\omega}=10 \text{ and } \frac{\sum_{m=1}^{N}D_m^{(4)}}{\omega}=5.$$

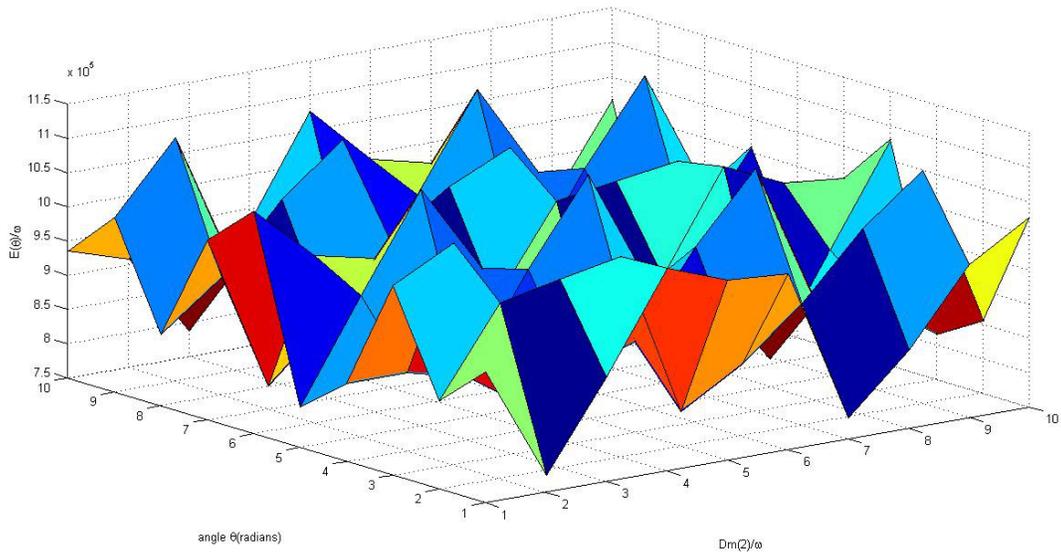

Figure 1: 3-D graph of $\frac{E(\theta)}{\omega}$ versus $\frac{\sum_{m=1}^{N}D_m^{(2)}}{\omega}$ and θ

By using the $\dfrac{\sum_{m=1}^{N} D_m^{(2)}}{\omega}$ values for each of the maxima and minima of the 3D graph, the 2-D graphs of $\dfrac{E(\theta)}{\omega}$ versus θ were plotted in order to find easy and hard directions of magnetization.

One of those graphs is shown in figure 2 for $\dfrac{\sum_{m=1}^{N} D_m^{(2)}}{\omega} = 44$ (one of the maximum energy points) and another graph is shown in figure 3 for $\dfrac{\sum_{m=1}^{N} D_m^{(2)}}{\omega} = 55$ (one of the minimum energy points).

According to figure 2, the hard directions are 0.2, 3, 6.4 radians ----etc for given values of energy parameters. The easy directions are 1.6, 4.5, 7.9 radians ------etc from figure 3.

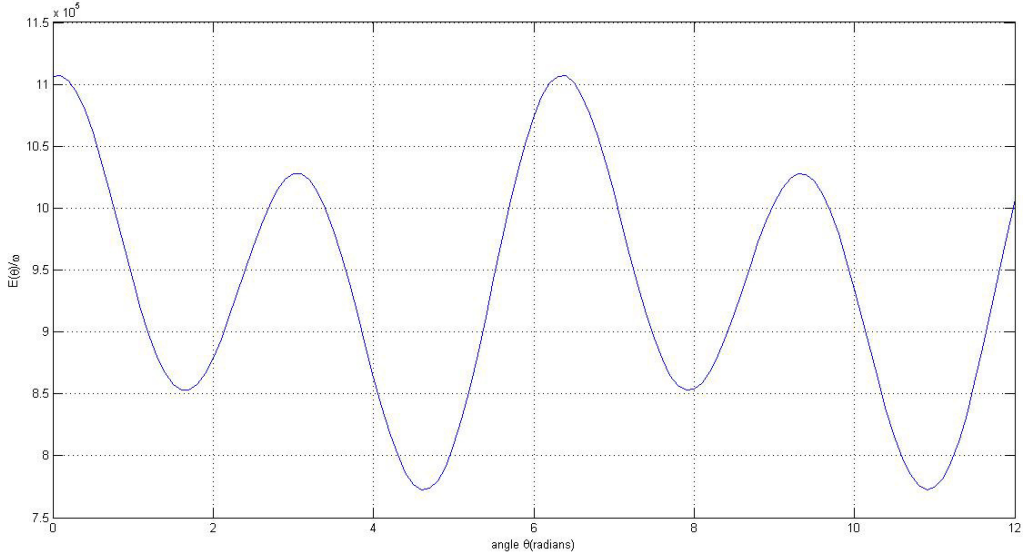

Figure 2: The graph of $\dfrac{E(\theta)}{\omega}$ versus θ at $\dfrac{\sum_{m=1}^{N} D_m^{(2)}}{\omega} = 44$ (one of the maximum energy points in 3-D plot).

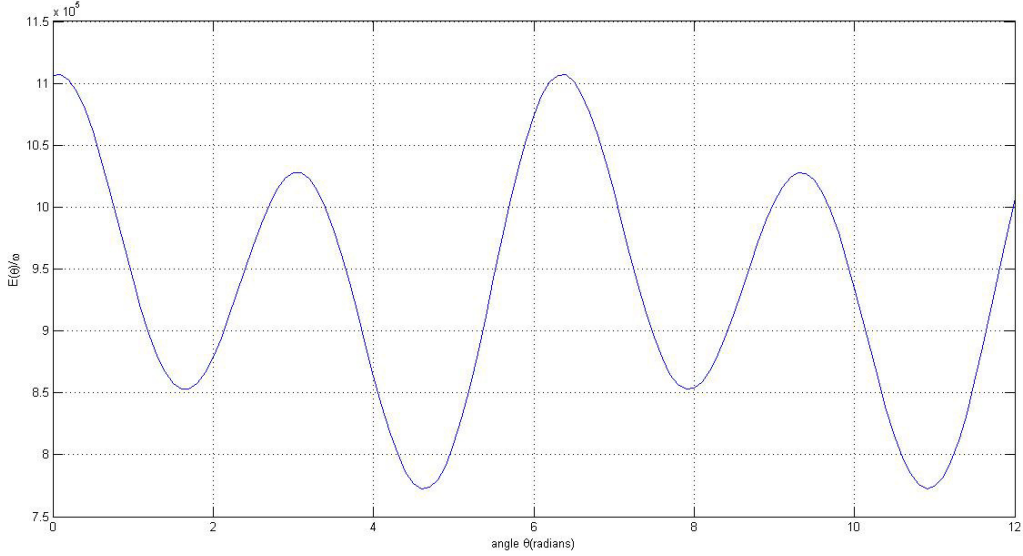

Figure 3: The graph of $\frac{E(\theta)}{\omega}$ versus $\theta$ at $\frac{\sum_{m=1}^{N} D_m^{(2)}}{\omega} = 55$ (one of the minimum energy points in 3-D plot).

## 4. Conclusion:

The total magnetic energy obtained using modified Heisenberg Hamiltonian varies periodically with energy parameters. The magnetic easy and hard directions were determined using 2-D and 3-D plots. The variation of energy is somewhat similar for $\frac{\sum_{m=1}^{N} D_m^{(2)}}{\omega}$ and $\frac{\sum_{m=1}^{N} D_m^{(4)}}{\omega}$. Higher variations could be observed for larger values of $\frac{H_{in}}{\omega}, \frac{H_{out}}{\omega}$ and $\frac{K_s}{\omega}$. When the number of layers or $\frac{J}{\omega}$ increases, the peak value of energy also increases. Changing $\frac{\sum_{m=1}^{N} D_m^{(2)}}{\omega}$, $\frac{\sum_{m=1}^{N} D_m^{(4)}}{\omega}$, $\frac{J}{\omega}, \frac{K_s}{\omega}$ or N does not change the easy axis or hard axis very much. But the angle of the hard axis increases with $\frac{H_{in}}{\omega}$, and the angle of the easy axis decreases with $\frac{H_{out}}{\omega}$.